\documentclass{svjour3}                     
\smartqed  
\usepackage{cite}
\usepackage{graphicx}
 \journalname{}
\begin{document}

\title{Thermodynamics and Phase transition from regular Bardeen black hole}


\author{Mahamat Saleh         \and
       Bouetou Bouetou Thomas \and
         Timoleon Crepin Kofane 
}


\institute{Mahamat Saleh \and
           Bouetou B T \and T C Kofane \at
              Department of Physics, Faculty of Science, University of Yaounde I, P.O. Box. 812, Cameroon \\
           \and
           Mahamat Saleh
             \at
              Department of Physics, Higher Teachers' Training College, University of Maroua, P.O. Box. 55, Cameroon, \\\email{mahsaleh2000@yahoo.fr}
              \and
              Bouetou Bouetou Thomas
           \at
              National Advanced School of Engineering, University of Yaounde I, P.O. Box 8390, Cameroon, \\\email{tbouetou@yahoo.fr}
              \and
              Kofane Timoleon Crepin
              \at
              The Max Planck Institute for the Physics of Complex Systems,
N\"othnitzer Strasse 38, 01187 Dresden, Germany \\\email{tckofane@yahoo.com}
            }

\date{Received: date / Accepted: date}

\maketitle

\begin{abstract}
In this paper, thermodynamics and phase transition are investigated for the regular Bardeen black hole. Considering the metric of the Bardeen spacetime, we derived the Unruh-Verlinde temperature. Using the first law of thermodynamics, we derived the expression of the specific heat and plot its behavior. It results that the magnetic monopole charge $\beta$ reduces the temperature and induces a thermodynamics phase transition in the spacetime. Moreover, when increasing $\beta$, the transition point moves to higher entropy.

\keywords{Black hole \and Thermodynamics \and Phase transition}
\end{abstract}

\section{Introduction}
\label{intro}
Nowadays research in physics focused a lot attention on the evolution of the Universe and particularly on black holes and strange phenomena enclosed to it. Over the last four decades, many researches were done on black hole physics and concern quasinormal modes\cite{R9, R10, R11, R12, R12a, R13, R14, R14a, R15, R16, R17, MelMos, R18, R19, R20, R20a, R20b,mah,mah2}, energy distribution\cite{R1,R2,R3,R4,R5,R6,R7,cs,xulu2,rad, vag,vag2,lali,vag3,virb1, virb2, virb3, aguir,xul, rad1,rad2, rad3,xulubt,msbk}, radiation and thermodynamics\cite{haw, haw2, bek2, pete, fabnag, zhao2hu, gibhaw, carlip1, carlip2, stromva, prl80, pw2,wald, bacaha}. Hawking's discovery of thermal radiation from black holes was a complete surprise to most specialists. This discovery definitely adds thermodynamical aspects to black hole investigation.

As thermodynamical objects, black holes can be described by some thermodynamics quantities such as temperature, entropy, enthalpy, free energy and heat capacity. One of the most interesting phenomena in the study of thermodynamical systems is their phase transition. Thermal fluctuations can induce phase transitions in
which the zero-temperature degrees of freedom are reorganized into a qualitatively different form. There are many familiar examples such as the boiling of water, ferromagnetic transitions (magnetisation), phase transitions also occur in more exotic systems like spacetime geometry\cite{pht1, pht2, pht3, phtphd, pht4, pht5, pht6}. The phase transition point of a system is generally represented by a discontinuity in the variation of such thermodynamical quantity of the system. For a glassy transition for example, there is a discontinuity in the variation of the volume with the temperature at the zero temperature. It is well known that at that particular temperature, the ice-water transition held.

Although the phase transition of several systems, such as water, can be investigated precisely in laboratory, no laboratory is yet equipped with the tools necessary to probe the formation of a black hole in a phase transition of thermal spacetime, although such situations may have existed in the very early universe\cite{pht3}. The black hole phase transition can be studied theoretically in the light of the expression of its heat capacity.


In 1968, the first example of a black hole with regular non-singular geometry with an event horizon satisfying weak energy condition was constructed by Bardeen. The solution was obtained introducing an energy-momentum tensor interpreted as the gravitational field of some sort of non linear magnetic monopole charge $\beta$.

Recently, Huang \emph{et al.}\cite{huang} considered the spherically symmetric regular black hole solution obtained by Bardeen to investigate absorption cross section and Hawking radiation. They show that the magnetic monopole charge, $\beta$, increases the absorption cross section and the power emission spectra of Hawking radiation but decreases the absorption probability and the luminosity. Meitei \emph{et al.}\cite{meitei} investigate phase transition in the Reissner-Nordstr\"om black hole. Bouetou \emph{et al.}\cite{bouetou} investigated thermodynamics and phase transition  of the Reissner-Nordstr\"om black hole surrounded by quintessence. In this paper, thermodynamics and phase transition are investigated for the regular Bardeen black hole.

The paper is organized as follows. In section ~\ref{sec:1}, we derive the Unruh-Verlinde temperature and the Hawking temperature for the regular Bardeen black hole. In section ~\ref{sec:2}, we derive the specific heat and investigate the thermodynamics phase transition in the black hole. The last section is devoted to a conclusion.

\section{Unruh-Verlinde and Hawking temperatures for the regular Bardeen black hole}
\label{sec:1}
The spherically symmetric Bardeen regular
black hole metric is given by
\begin{equation}\label{e1}
    ds^2=-f(r)dt^2+f^{-1}(r)dr^2+r^2d\theta^2+r^2\sin^2\theta d\varphi^2,
\end{equation}
the lapse function $f(r)=1-\frac{2M(r)}{r}$ depends on the specific form of underlying matter. With the
following particular value of the mass function,
\begin{equation}\label{e2}
    M(r)=\frac{mr^3}{(r^2+\beta^2)^{3/2}},
\end{equation}
where $\beta$ is the monopole charge of a self-gravitating magnetic field described by a nonlinear electrodynamics source, and $m$ is the mass of the black hole, the metric (\ref{e1}) reduces to the Bardeen regular black
hole metric\cite{bard,huang,sharif,sharif1}.

The temperature of the black hole at a given radius $r$ is given by the Unruh-Verlinde matching\cite{Verlinde, Unruh}. The Unruh-Verlinde temperature is given by\cite{Konoplya, Liuwangwei,hanlan}
\begin{equation}\label{at1}
    T=\frac{\hbar}{2\pi}e^\phi n^\alpha\nabla_\alpha\phi,
\end{equation}
where $n_\alpha$ is a unit vector, that is normal to the holographic
screen, $e^\phi$ is the red-shift factor and $\phi$ the generalized form of the Newtonian potential given by
\begin{equation}\label{at2}
    \phi=\frac{1}{2}\log(-g^{\mu\nu}\xi_\mu\xi_\nu),
\end{equation}
with $g^{\mu\nu}$ the background metric and $\xi_\mu$ the Killing time-like vector.
The Killing vector for our spherically symmetric spacetime is
\begin{equation}\label{at3}
    \xi_\mu=\left(-f(r), 0, 0, 0\right).
\end{equation}
Substituting Eq. (\ref{at3}) into Eq. (\ref{at2}), we can obtain the following expression for the potential
\begin{equation}\label{at4}
    \phi=\frac{1}{2}\log\left(f(r)\right)
\end{equation}
and the expression of the  Unruh-Verlinde temperature is given by
\begin{equation}\label{e3}
    T=\frac{\hbar}{4\pi}|f'(r)|.
\end{equation}
Substituting Eqs. (\ref{e1}) and (\ref{e2}) into (\ref{e3}), the Unruh-Verlinde temperature for the regular Bardeen black hole is
\begin{equation}\label{e4}
    T=\frac{mr(r^2-2\beta^2)}{2\pi(r^2+\beta^2)^{5/2}}.
\end{equation}
Setting $\beta=0$, the temperature becomes $T=\frac{m}{2\pi r^2}$, which is the Unruh-Verlinde temperature for the Schwarzschild black hole.

The event horizon of the black hole is given by
\begin{equation}\label{e5}
    f(r_h)=1-\frac{2mr_h^2}{(r_h^2+\beta^2)^{3/2}}=0.
\end{equation}
The entropy of the black hole is given by the area law
\begin{equation}\label{e6}
    S=\frac{A}{4}=\pi r_h^2.
\end{equation}
The mass of the black hole can then be expressed as function of the entropy as follows
\begin{equation}\label{e7}
    m=\frac{1}{2}\sqrt{\frac{S}{\pi}}\left(1+\frac{\pi\beta^2}{S}\right)^{3/2}.
\end{equation}

Applying the first law of thermodynamics, the Hawking temperature (temperature at the horizon) is
\begin{equation}\label{e8}
    T_H=\frac{1}{4\sqrt{\pi S}}\left(1+\frac{\pi\beta^2}{S}\right)^{1/2}\left(1-\frac{2\pi\beta^2}{S}\right).
\end{equation}
This expression of the temperature can be used to derive other thermodynamics quantities of the black hole.

\begin{figure}[h!]
  \includegraphics[width=8cm]{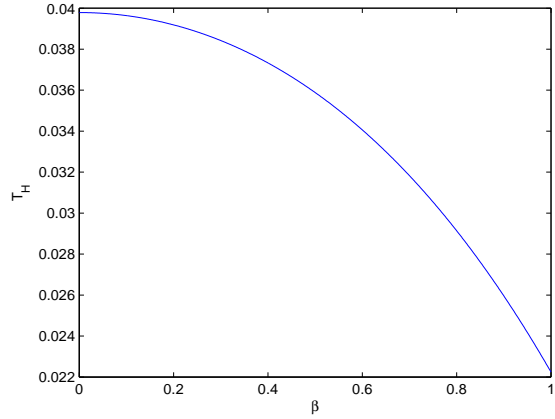}\\
  \caption{Behavior of the Hawking temperature versus $\beta$.}\label{ht}
\end{figure}

The behavior of the temperature is shown on Fig. \ref{ht}.

Through this figure, we can see that the temperature decreases when increasing $\beta$.

\section{Specific heat and phase transition}
\label{sec:2}
\begin{figure}[h!]
  \includegraphics[width=10cm]{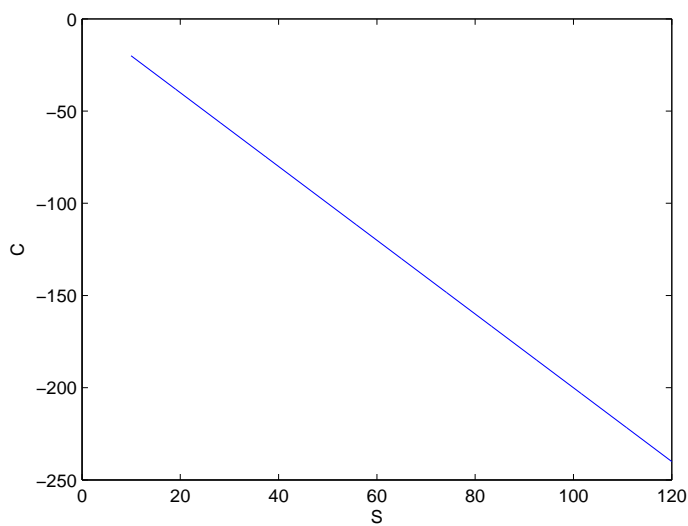}\\
  \caption{Behavior of the specific heat $C$ versus $S$ for  $\beta=0$}\label{hcb}
\end{figure}

The internal energy of the black hole of mass $m$ can be expressed by the Einstein formula
$E=mc^2$. Considering this expression and taking into account the first law of thermodynamics, the temperature of the black hole can be derived as:
\begin{equation}\label{sh1}
    T=\frac{\partial E}{\partial S},
\end{equation}
with $S$ the entropy of the black hole. The specific heat of the black hole is then given by
\begin{equation}\label{sh2}
    C=T\frac{\partial S}{\partial T}.
\end{equation}
Using the above equations, the specific heat can then be expressed as:
\begin{equation}\label{sh3}
    C=-2S\frac{(S+\pi\beta^2)(S-2\pi\beta^2)}{S^2-4\pi\beta^2S-8\pi^2\beta^4}.
\end{equation}
For $\beta=0$, this expression reduces to $C=-2S$ which corresponds to the expression for the Schwarzschild black hole. The specific heat is negative (see Fig. \ref{hcb}) showing that the black hole is thermodynamically unstable.

\begin{figure}[h!]
  \includegraphics[width=10cm]{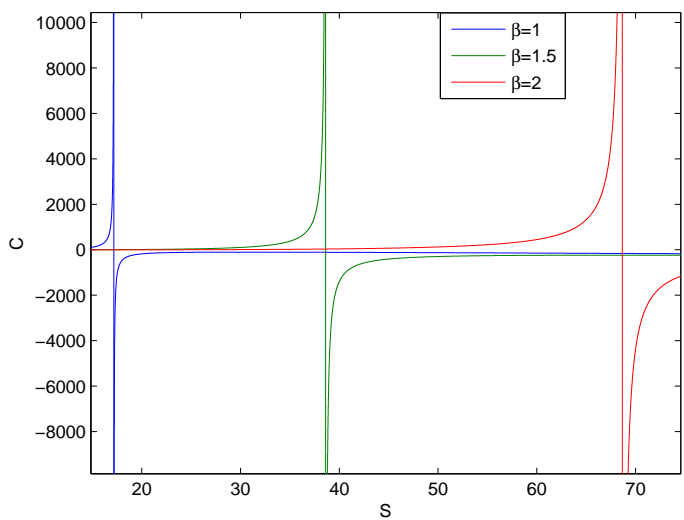}\\
  \caption{Behavior of the specific heat $C$ versus $S$ for different values of $\beta$}\label{pht1}
\end{figure}

\begin{figure}[h!]
  \includegraphics[width=10cm]{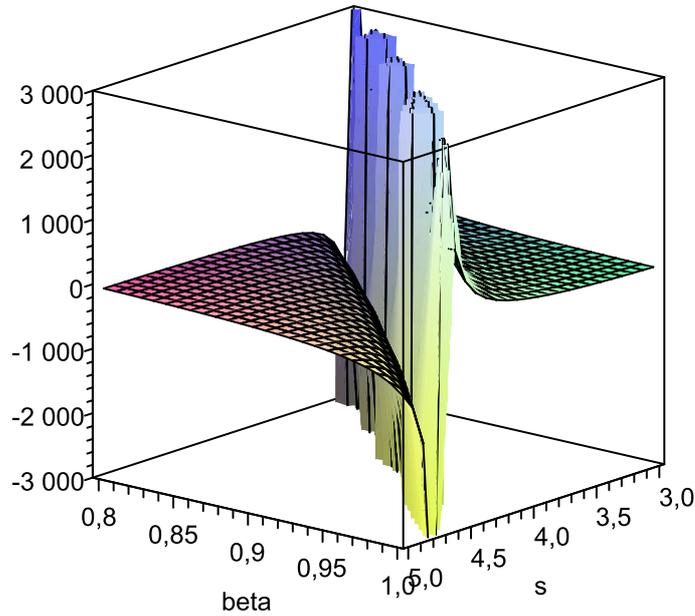}\\
  \caption{Behavior of the specific heat $C$ versus $\beta$ and $S$}\label{pht2}
\end{figure}

For $\beta\ne0$, we plot the behavior of the specific heat of the black hole when increasing $S$ for different values of $\beta$. This is shown on Figs.\ref{pht1} and \ref{pht2}.

Through these figures, we can remark that when taking into account such positive value of $\beta$, it moves the specific heat to positive value for lower entropies showing that the black hole is at stable phase. Moreover, when increasing the entropy $S$, it occurs a discontinuity point where the specific heat changes from positive to negative values showing that the black hole passes from stable to unstable phase. When increasing $\beta$, the transition point moves to higher entropy.

\section{Conclusion}
\label{conc}
Thermodynamics behavior of the regular Bardeen black hole is investigated. From the metric of the black hole, the Unruh-Verlinde temperature is expressed. The Hawking temperature of the black hole and the specific heat are also derived using the laws of black holes thermodynamics. The behavior of the Hawking temperature plotted on Fig. \ref{ht} shows that the monopole charge $\beta$ decreases the temperature of the black hole. Through the behavior of the specific heat of the black hole plotted on Figs. \ref{pht1} and \ref{pht2}, we can see that the monopole charge $\beta$ stabilizes thermodynamically the black hole and when increasing the entropy of the black hole, it occurs a transition point where the black hole moves from stable thermodynamic phase to unstable one. Moreover, the transition point moves to higher entropies when increasing $\beta$.

\end{document}